\documentclass{article}

\usepackage{amsmath,amssymb,amsfonts,dsfont}
\usepackage{diagbox}
\usepackage[backend=biber,style=vancouver]{biblatex}
\usepackage{authblk} 
\usepackage{graphicx}
\usepackage[a4paper, margin=2.54cm]{geometry}
\usepackage{lmodern}

\bibliography{GroverAlgorithm.bib}

\title{Improvement of performance of Grover's algorithm on three generations of Heron family IBM QPUs without and with topological dynamical decoupling}

\author[1]{Tihomir G. Tenev} 
\author[2]{Nayden P. Nedev} 
\author[2]{Nikolay V. Vitanov}

\affil[1]{Georgi Nadjakov Institute of Solid State Physics, Bulgarian Academy of Sciences, 72, Tzarigradsko Chaussee Blvd., 1784 Sofia, Bulgaria}

\affil[2]{Center for Quantum Technologies, Department of Physics, St. Kliment Ohridski University of Sofia, 5 James Bourchier blvd, 1164 Sofia, Bulgaria}

\begin{document}

\maketitle

\begin{abstract}
We investigate the performance of Grover's algorithm on three different generations of IBM Heron QPUs. On Heron family of IBM QPUs the success probabilities for three, four and five qubits without dynamical decoupling is better than results reported for previous generations of QPUs. The success probability as function of number of iterations of Grover operator is considered. A study of the improvement of results of Grover's algorithm for five qubit case with the help of topological dynamical decoupling is considered. For a six qubit case on Heron r3 QPU a clear result for finding the sought-after bitstring is reported for theoretically suboptimal number of iterations of Grover operator with the help of dynamical decoupling.
\end{abstract}

\section{Introduction}
The classical algorithms for searching of unsorted list of $N$ elements have\cite{watrousUnderstandingQuantumInformation2025} computational complexity $O(N)$. The Grover's algorithm\cite{groverQuantumMechanicsHelps1997} is one of the first algorithms of quantum information. It offers ability to find an element in unsorted list with the order of $\sqrt{N}$ steps, that is it has computational complexity $O(\sqrt{N})$. In principal this can offer quadratic speedup over classical algorithms. 

Early attempts have been made to realize Grover's algorithm in various physical systems\cite{vandersypenImplementationThreequantumbitSearch2000,anwarImplementingGroversQuantum2004,brickmanImplementationGroversQuantum2005,yangImplementationThreequbitGrover2007,liuFirstExperimentalDemonstration2015,figgattComplete3QubitGrover2017} most of which have been limited to up to three qubit register size. 
With the availability of quantum computers over the cloud a number of attempts have been made to realize Grover's algorithm on them\cite{mandviwallaImplementingGroversAlgorithm2018,jingleDesignGroversAlgorithm2021,zhangImplementationEfficientQuantum2021,zhangQuantumSearchNoisy2022,elkaderiPerformanceUncodedImplementation2023,
parkQuantumMultiprogrammingGrovers2023,pokharelBetterthanclassicalGroverSearch2024,abughanemCharacterizingGroverSearch2025}.  There are also recent studies reporing realization of Grover's algorithm in modalities different from the mainstream ones\cite{gustianiBlindThreequbitExact2021,zhangSimulationGroverSearch2023,royTwoQutritQuantumAlgorithms2023,simonettiIMPLEMENTINGTHREEQUBITGROVERS2025,thorvaldsonGroversAlgorithmFourqubit2025}. Due to combination of limited number of qubits in quantum hardware in earlier studies or gate fidelity and consequent limits on quantum circuit depths most studies have been focused on the range of between two and five qubits for the length of searched bitstrings. Some of the reported five qubit studies\cite{zhangQuantumSearchNoisy2022, pokharelBetterthanclassicalGroverSearch2024} have been performed with reduced number of iterations of Grover operator.

In this study we focus on investigating the performance of Grover's algorithm on IBM Heron family of superconducting QPUs for problem sizes of between three and six qubits. We report better success probabilities without error correction for three, four and five qubits on IBM Heron family of QPUs compared with studies on previous generations of superconducting QPUs. We investigate the improvement of performance of Grover's algorithm for the five qubit case with the help of the recently proposed topological dynamical decoupling\cite{nedevTopologicalDynamicalDecoupling2025a} and compare it to the well established CPMG and XY4 dynamical decoupling sequences. We investigate the performance of Grover's algorithm for the six qubit case on the considered QPUs for different iterations of Grover operator enhanced with selected dynamical decoupling sequences.. These runs show above average success probability for six qubit problem size for up to three iterations of Grover operator on the latest generation of Heron family QPUs when selected dynamical decoupling sequences are used in the quantum circuits. 

\section{Methods}
\subsection{Brief description of Grover's algorithm}
The Grover's algorithm requires a register of $n$ qubits to search through $N=2^n$ bitstrings of length $n$. It's first step\cite{watrousUnderstandingQuantumInformation2025} is to initialize the quantum register in equal superposition $\vert u \rangle = \hat{H}^{\otimes n}\vert 0^n\rangle = \frac{1}{\sqrt{N}}\sum\vert x\rangle$ of all possible states by application of Hadamard gates $\hat{H}$ on each qubit. Then the Grover operator $\hat{G}$ is applied certain number of times depending on the size $N$ of the unsorted list in which the search is performed and the number of sought after solutions $s$.  The Grover operator $\hat{G}=\hat{D}\hat{O}_{t}$ is made of two operators - the oracle operator $\hat{O}_{t}$ which construction depends on the target state $\vert t \rangle$ and the operator $\hat{D}$ sometimes referred to as a diffusion operator or amplitude amplification step. The oracle operator $\hat{O}$ marks the target state by introducing a phase factor in front of it 
\begin{equation}
\hat{O}_{t}\vert x\rangle = (-1)^{f(x)}\vert x\rangle = \begin{cases}\begin{array}{llll} 
 -\vert x\rangle & , & \text{if} & x=t, \\
  \vert x\rangle & ,  & \text{if} & x\neq t. 
\end{array}
\end{cases}  
\end{equation}
as the function $f(x)$ is chosen such that $f(x)\vert_{x=t}=1$ and $f(x)\vert_{x\neq t}=0$.
This can also be thought as a reflection of the state acted upon around the axis in the Hilbert space of the problem defined by equal superposition of all non-solutions.

Then the diffusion operator or amplitude amplification step
\begin{equation}
\hat{D} = \hat{H}^{\otimes n} (2\vert 0^n\rangle \langle 0^n\vert -  \hat{\mathds{1}}) \hat{H}^{\otimes n} = 2\vert u\rangle\langle u \vert - \hat{\mathds{1}}
\end{equation}
is applied, which can be thought of as a reflection of the state acted upon around the axis in the Hilbert space of the problem defined by the average superposition $\vert u \rangle$.
The combination of two reflections is equivalent to rotation by angle $2\theta$ in the Hilbert space of the problem, where for single sought after element $\theta=\arcsin\left(\sqrt{\frac{1}{N}}\right)$. The optimal number of steps for repeating the Grover operator is obtained by rounding $\frac{\pi}{4\theta}$ to the nearest smallest integer.

\subsection{Implementation}
We use IBM Qiskit to implement\cite{ibmGroversAlgorithm} the Grover oracle and Grover operator. The Grover oracle is implemented by sandwiching multi-control Z gate between X gates applied to $0$-valued qubits in the target state. The Grover operator is constructed by supplying the constructed Grover oracle to the IBM Qiskit \texttt{grover\_operator()} function\cite{ibmGroversAlgorithm}. The implementation of the diffusion operator $\hat{D}$ that realizes the amplitude amplification step also requires the multi-control Z gate sandwiched between X gates. Therefore for one iteration of the Grover operator $\hat{G}$ two implementations of multi-control Z gate is required. Since the multi-control Z gate is not from the native set of QPUs it has to be implemented in terms of native 1Q and 2Q gates. The number of 1Q and 2Q gates required to implement the multi-control Z gate depends on the size of the quantum register. In practice this decomposition is handled by the IBM Qiskit transpiler. 

We have used qiskit version 2.1.0 and qiskit\_ibm\_runtime version 0.40.1. We have used optimization level 3 for the transpiler for all the runs in this study. For the sake of reproducibility the seed to the transpiler has been fixed to the same value of 1234 for all runs of this study. The same set of initial qubits have been used for all the runs and respective figures of the five qubit case on the considered QPUs. All runs have been executed with 10000 shots.

\subsection{Description of the considered dynamical decoupling sequences}
Previous studies\cite{pokharelBetterthanclassicalGroverSearch2024} have investigated the effects of measurement error mitigation and various dynamical decoupling (DD) sequences on the success probability of Grover's algorithm for three, four and five qubit registers. It has been found that measurement error mitigation has marginal effect on the improvement of the success probability while various dynamical decoupling sequences have significant effect. For this reason first we do not investigate the effects of measurement error mitigation. In this work we focus on the study of the effect on Grover's algorithm success probability of recently proposed dynamical decoupling family of sequences $Tn$ called topological dynamical decoupling\cite{nedevTopologicalDynamicalDecoupling2025a} and compared them to the well established CPMG and XY4.

The well known CPMG sequence proposed by Carr and Purcell \cite{carrEffectsDiffusionFree1954a} and Meiboom and Gill \cite{meiboomModifiedSpinEchoMethod1958a} is defined as
\begin{equation}
CPMG \equiv  X - X 
\end{equation}  

The XY4 sequence that we use has the usual definition 
\begin{equation}
XY4 \equiv  X - Y - X-  Y  
\end{equation} 

The $Tn$ sequences\cite{nedevTopologicalDynamicalDecoupling2025a} are subdivided into two sub-classes $T4l$ and $T2m$.  The $T4l$ sequences have number of pulses which are divisible by 4. Their structure is
\begin{equation}
T4l = (r_l)_0(r_l^{-1})_0(r_l)_{\pi}(r_l^{-1})_{\pi}
\end{equation}
where
\begin{subequations}
\begin{align}
(r_l)_0 &= (\phi_1, \phi_2,...,\phi_{l-1},\phi_l) , \\
(r_l^{-1})_0 &= (\phi_l, \phi_{l-1},...,\phi_2,\phi_1) , \\
(r_l)_{\pi} &= (\phi_{1}+\pi, \phi_{2} + \pi,...,\phi_{l-1} + \pi, \phi_l + \pi) , \\
(r_l^{-1})_{\pi} &= (\phi_l + \pi, \phi_{l-1} + \pi,...,\phi_2 + \pi, \phi_1 + \pi)
\end{align}
\end{subequations}
and $l=1,2,...$ . The sequences $T2m$, where $m=2j+1$ and $j=0,1,2,...$ are those that have number of pulses divisible by two but not by four and have the structure
\begin{equation}
T2m = (r_m)_0 (r_m)_{\pi} ,
\end{equation}
where 
\begin{subequations}
\begin{align}
(r_m)_0 & = (\phi_1, \phi_2,...,\phi_{m-1},\phi_m) , \\
(r_m)_{\pi} & = (\phi_1 + \pi, \phi_2 + \pi,...,\phi_{m-1}+\pi, \phi_m + \pi  )
\end{align}
\end{subequations}
The formula for the phases of both sequences is
\begin{equation}
\phi_k = \frac{(k-1)(n/2-k)}{n/2}\pi ,
\end{equation}
where $n$ denotes the total number of pulses in the sequence and $k=1,2,...,n$.

We implement the topological dynamically decoupled sequences by combination\cite{krantzQuantumEngineersGuide2019a,vezvaeeVirtualGatesSymmetric2025} of X gate sandwiched between two $R_z$ gates with appropriately chosen phases $R_z(-\phi)XR_z(\phi)$ separated by time intervals. The dynamically decoupling (DD) sequences are inserted into the circuit with the help of PassManager with  \texttt{ALAPScheduleAnalysis} and  \texttt{PadDynamicalDecoupling} passes. Furthermore the  \texttt{BasisTranslator} pass is used to translate any nonnative gates to gates native for the particular QPU. The DD sequence fill the available idle spaces of the circuit with idle time between the pulses determined by algorithms  based on number of pulses and available time. This leads to different idle times between pulses for different idle intervals.

\section{Results and Discussion}
\subsection{Grover's algorithm for  problem size of three to five qubits}
\begin{figure}[ht]
    \centering
    \includegraphics{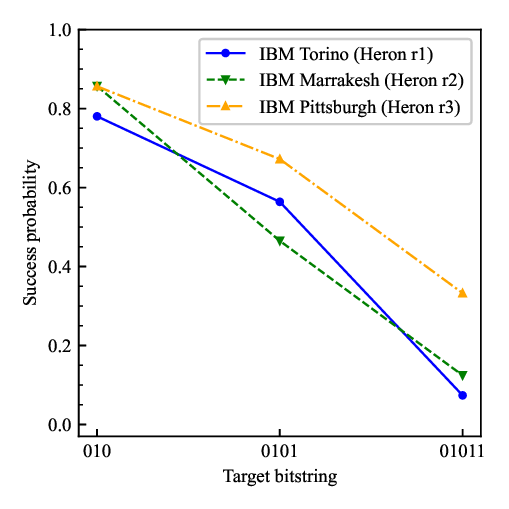}   
    \caption{(Color online) Success probabilities for Grover's algorithm for three to five qubits on different generations of IBM QPUs without any form of error mitigation. The reported values of success probabilities for the three, four and five qubit cases are higher than these reported in previous studies of superconducting QPUs. There is clear trend of improvement of performance for the success probabilities with newer generations of the Heron family QPUs.}
    \label{fig:Grover3to6qubits}
\end{figure}

On figure~\ref{fig:Grover3to6qubits} results from runs without error correction on  Torino (Heron r1) and Marrakesh (Heron r2) and Pittsburgh (Heron r3) are presented for three to five qubits for target bitstrings "010", "0101", "01011". The target states are chosen with balanced number of 0-s and 1-s to account for possible variations of results if more 1-s are chosen in the bitstring. This is so because 1-a are implemented as excited states which have some coherence time and in general may decay. The runs on Torino  (Heron r1) were performed on  2026-03-06, the ones on Marrakesh  (Heron r2) on 2025-03-12 and the ones on Pittsburgh  (Heron r3) on 2026-03-09. They were performed on different dates as we have recorded calibration data T1, T2, readout errors and 2Q gate errors of the chosen qubits for over two weeks and the runs were executed when most or all of the mean and min values of T1, T2 and mean and max values of readout errors and 2Q errors of the chosen qubits were below or above average as appropriate for the considered metric for the monitored period. The calibration metrics used to choose the runtime are shown in table~\ref{table:Grover3QubitProperties} for three qubits, table~\ref{table:Grover4QubitProperties} for four qubits and table~\ref{table:Grover5QubitProperties} for five qubits. Every run has been performed with 10000 shots. 

For three qubits the success probabilities for Marrakesh (Heron r2) and Pittsburgh (Heron r3) are comparable to each other and higher than the result for Torino (Heron r1). For four qubits the success probability for Torino (Heron r1) is higher than that of Marrakesh (Heron r2) and both are smaller than the result on Pittsburgh (Heron r3).  For five qubits the success probability on Marrakesh (Heron r2) is slightly better than that of Torino (Heron r1) while the success probability on Pittsburgh (Heron r3) is significantly higher than that on the previous two generations of Heron QPUs. 

The results on all three considered QPUs without error correction for three and four qubits show higher success probability than error corrected ones reported\cite{pokharelBetterthanclassicalGroverSearch2024} on earlier generation of IBM QPUs. For five qubits the results without error mitigation for Torino (Heron r1) and Marrakesh (Heron r2) are comparable to dynamically decoupling corrected results on previous generations of IBM QPUs\cite{pokharelBetterthanclassicalGroverSearch2024}. However the success probability of Pittsburgh (Heron r3) without error correction is larger than reported results with dynamical decoupling for previous generations of IBM QPUs. The improvement of qubit properties with newer generations of QPU as shown in table~\ref{table:Grover3QubitProperties} for three qubits, table~\ref{table:Grover4QubitProperties} for four qubits and  table~\ref{table:Grover5QubitProperties} for five qubits considerably improves the performance of Grover's algorithm for the cases of three, four and five qubits.

\subsection{Grover's algorithm for five qubits with different number of iterations}
\begin{figure}[ht]
    \centering
    \includegraphics{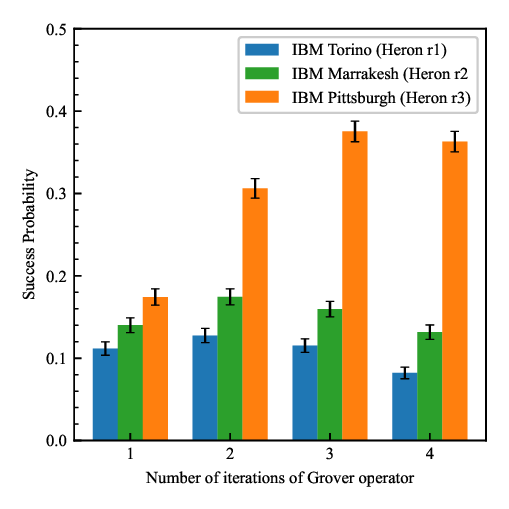}   
    \caption{(Color online) Success probabilities for Grover's algorithm for five qubits target state "01011" with different number of iterations on different generations of IBM QPUs. The error bars represent 99\% binomial confidence intervals.}
    \label{fig:Grover5qubitsDifferentIterations}
\end{figure}

In order to get better insight of the performance of Grover's algorithm for the five qubit case, runs were performed for theoretically suboptimal number of iterations of Grover operator and the results without dynamical decoupling are presented in figure~\ref{fig:Grover5qubitsDifferentIterations}. For the five qubit case the target state is chosen to be the same  $\vert01011\rangle$ balanced in terms of 0-s and 1-s states used in the runs from three to five qubits. The data for each iterations is obtained from runs with 10000 shots. The error bars represent 99 percents binomial confidence intervals.

Theoretically the highest success probability should be achieved for optimal number of iterations which for the five qubits case is four iterations. However practically figure~\ref{fig:Grover5qubitsDifferentIterations} shows that the highest success probability is achieved for smaller number of iterations and that the exact number of iterations can be QPU dependent. On Torino (Heron r1) the highest success probability is achieved for two iterations. On Marrakesh (Heron r2) the success probabilities for two iterations is also higher than the case with three iterations. On Pittsburgh (Heron r3) the highest success probability is achieved for three iterations and it is nearly 0.38. The exact number of iterations at which the maximum of success probability is achieved can be looked as a trade-off between the theoretically optimal performance and the rate of accumulation of decoherence due two 2Q gate errors and idle time decoherence. The shift of the peak performance with number of iterations with newer generations of QPUs can be accounted by the decrease of 2Q gate errors and their accumulation and the increase of T1 and T2 of qubits. The number of 2Q gate operations for different number of iterations is presented in table~\ref{table:Grover4IterationsNumberOf2QOperations5Qubit} and is the same for Torino (Heron r1), Marrakesh (Heron r2) and Pittsburgh (Heron r3).

\begin{table}[h]
\caption{Number of 2Q operations used in the calculations of intermediate number of iterations for Grover's algorithm implemented for five qubits.}
\label{table:Grover4IterationsNumberOf2QOperations5Qubit}
\begin{center}
\begin{tabular}{l c c c c}
\hline
\diagbox{QC}{Iterations}  & 1  & 2   & 3  & 4    \\ 
\hline
Torino (Heron r1) & 127  & 263  & 402 & 538   \\
Marrakesh (Heron r2) & 127  & 263  & 402 & 538   \\
Pittsburgh (Heron r3) & 127  & 263  & 402 & 538   \\
\hline
\end{tabular}
\end{center}
\end{table}

\subsection{Comparison between selected classical dynamical decoupling sequences and topological dynamical decoupling with up to twelve pulses for the five qubit case}

The properties of the topological dynamical decoupling sequences are examined on single-qubit system in Ref.~\cite{nedevTopologicalDynamicalDecoupling2025a}. The robustness against variations in detuning and pulse area were examined. Here we study the practical application of the $Tn$ sequences on five qubit case of Grover's algorithm. Results from Ref.~\cite{pokharelBetterthanclassicalGroverSearch2024} show that in general dynamical decoupling sequences with smaller number of pulses show better improvement of performance of the success probability compared with longer sequences. Therefore we focus on T2 to T12 sequences and compare them to the well known CPMG and XY4 dynamical decoupling sequences for reference.

\begin{figure*}[ht]
    \centering
    \includegraphics[width=\linewidth]{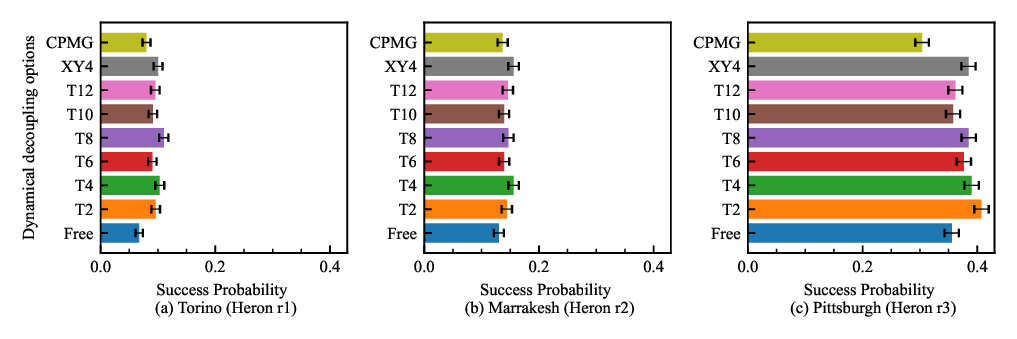}   
    \caption{(Color online) Success probabilities for Grover's algorithm with different dynamical decoupling options for five qubits on three generations of IBM Heron family QPUs. The error bars represent 99\%  confidence intervals.}
    \label{fig:Grover5qubitsDD}
\end{figure*}

On figure~\ref{fig:Grover5qubitsDD} are shown the results for five qubit case run on Torino (Heron r1),  Marrakesh (Heron r2) and Pittsburgh (Heron r3) from runs with 10000 shots for the considered dynamical decoupling sequences. The error bars represent 99 percents binomial confidence intervals. On Torino (Heron r1)  CPMG offers the smallest enhancement in success probability. Within the confidence intervals T2, T4, T6, T10 and T12 offer comparable enhancement to XY4 sequence, while T8 is the sequence with best performance. On Torino (Heron r1) the enhancement offered by the T8 sequence compared to the free case is around 30\%. On Marrakesh (Heron r2) the best performance is achieved by XY4 with T4 comparable to it within the 99\% confidence interval. The other Tn sequences offer slightly less enhancement compared to the best performing XY4 and T4, but still comparable to them within the 99\% confidence intervals, although their value is closer to the enhancement offered by CPMG sequence. On Pittsburgh (Heron r3) again CMPG has the worst performance and it actually decreases the success probability compared to the free case. On this QPU T10 and T12 offer worse enhancement compared to XY4, while T4, T6 and T8 offer comparable enhancement to XY4 and T2 offers slightly better enhancement of success probability compared to XY4. On Pittsburgh (Heron r3)  the enhancement offered by T2 compared to the free case is around 14\%.

Two trends are noticeable for the Tn sequences presented on figure~\ref{fig:Grover5qubitsDD}. First there seems to be oscillatory behavior in the success probability with respect to number of pulses. Second it seems that Tn sequences with smaller number of pulses show better enhancement compared to the T10 and T12 sequences. 

The absolute values of the success probabilities for the three considered QPUs depend on the hardware properties and used qubit calibrations. Clearly there is a trend of significant improvement of the both free and dynamically decoupling protected success probability with newer generations of QPUs. The calibrations that are used in figure~\ref{fig:Grover5qubitsDD} are reported in table~\ref{table:Grover5QubitProperties}. It is notable that the success probability for the case without dynamical decoupling for Pittsburgh (Heron r3) reaches the value 0.35 for the used calibration for the five qubit case which is almost twice as better as the reported values\cite{pokharelBetterthanclassicalGroverSearch2024} for dynamically decoupling protected runs on previous generations of QPUs. On Pittsburgh (Heron r3) with the corrections of the T2 sequence which has the best performance the success probability reaches 0.4 .

In an attempt to understand these results first we note that in Ref.~\cite{ezzellDynamicalDecouplingSuperconducting2023} it has been found that for a single qubit system by optimizing the interval between pulses in CMPG or XY4 sequences they can be made to match the performance of more advanced dynamical decoupling sequences like UR\cite{genovArbitrarilyAccuratePulse2017} and QDD\cite{westHighFidelityQuantum2010}.
Second we note that in the actual implementation of the $Tn$ sequences in Grover's algorithm quantum circuit the time delays between pulses in a sequence vary in different idle intervals of the same circuit. This variation depends on the available idle time in the interval. Therefore in the actual implementation of the dynamical decoupling sequences in Grover's algorithm circuit we use an ensemble of a particular kind of dynamical decoupling sequence with some distribution of the delays between the pulses that depends on circuit details and chosen qubit connectivity.

\begin{figure*}[ht]
    \centering
    \includegraphics[width=\linewidth]{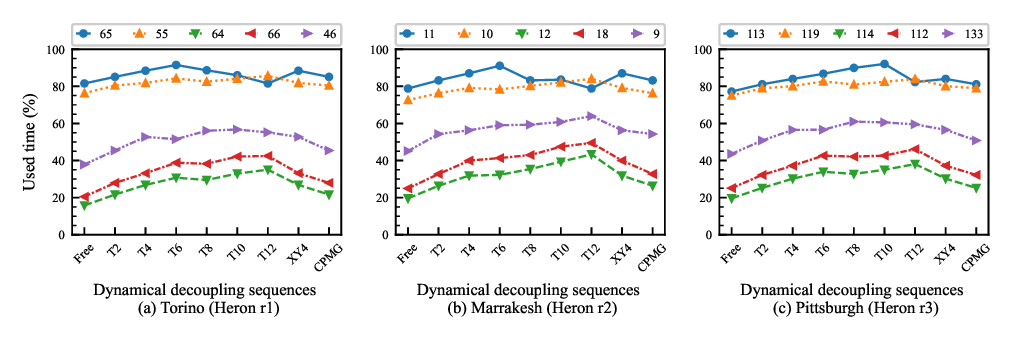}   
    \caption{(Color online)  Used time in percent of total circuit duration per qubit for Grover's algorithm with different dynamical decoupling options for five qubits on three generations of IBM Heron family QPUs. The legends show qubit numbers that were used in the implementation of the circuit. The circuit implemented on Torino has total duration of  $5.77\times10^{-5}$ s, on Marrakesh has total duration of $5.282\times10^{-5}$ s and on Pittsburgh has total duration of $6.022\times10^{-5}$ s.}
    \label{fig:Grover5qubitsDDUsedTime}
\end{figure*}

\begin{figure*}[ht]
    \centering
    \includegraphics[width=\linewidth]{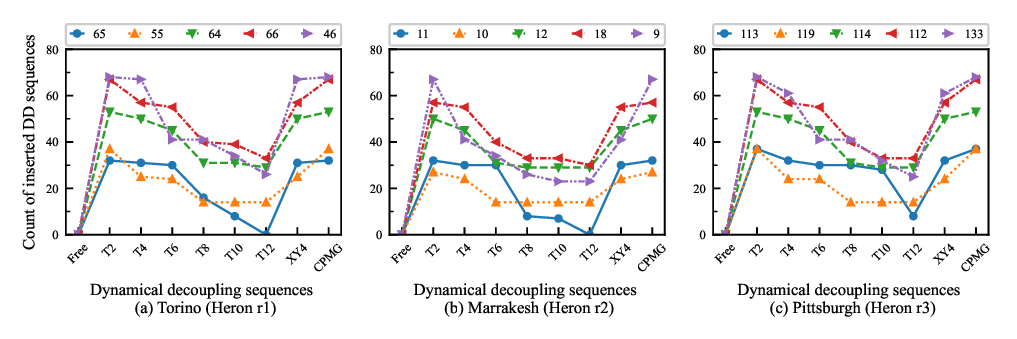}   
    \caption{(Color online)  Number of inserted DD sequences per qubit for Grover's algorithm with different dynamical decoupling options for five qubits on three generations of IBM Heron family QPUs. The legends shows qubit numbers that were used in the implementation of the circuit.}
    \label{fig:Grover5QubitsNumberInsertedDDSequences}
\end{figure*}

To gain further insight into the working of the dynamically decoupled sequences we have plotted the time used by gate operations per qubit presented on figure~\ref{fig:Grover5qubitsDDUsedTime} and the number of inserted DD sequences on figure~\ref{fig:Grover5QubitsNumberInsertedDDSequences}. The circuit implemented on Torino has total duration of  $5.77\times10^{-5}$ s, on Marrakesh has total duration of $5.282\times10^{-5}$ s and on Pittsburgh has total duration of $6.022\times10^{-5}$ s. On figure~\ref{fig:Grover5qubitsDDUsedTime} it is evident that on all three QPUs, two qubits experience the highest load of operations, one qubit is busy with operations nearly half the time of the circuit duration and two qubits are experience lower loads. This can be attributed to the star topology of qubits chosen to implement the circuit for Grover's algorithm. This implies that fidelity of gates applied to the two most busy qubits will contribute more to the performance of Grover's circuit, while the DD sequences will have stronger effect for the less busy qubits. This is evident by the fact that the relative increase of used time with different DD sequences is higher for the three qubits with lower overall time usage.

Indeed when we examine the number of inserted DD sequences as shown on figure~\ref{fig:Grover5QubitsNumberInsertedDDSequences} we see that the less busy qubits experience higher number of inserted DD sequences. T2 and CPMG are made of the same amount of pulses, namely two, and therefore the amount of inserted DD sequences for T2 and CPMG are comparable for every qubit. Also T4 and XY4 are made both of four pulses and therefore the amount of inserted DD sequences of this type for every considered qubit are almost the same on all considered QPUs. When looking at results of the success probability on figure~\ref{fig:Grover5qubitsDD} we see that the T2 with the same amount of pulses as CPMG offers better performance than CPMG. In contrast T4 offers comparable performance to XY4  which both are made of 4 pulses. This may be due to the fact that XY4 uses rotation around two different axis, namely X and Y, while T4 uses rotation about only the X axes.

Figure~\ref{fig:Grover5QubitsNumberInsertedDDSequences} shows another clear trend that the amount of inserted Tn sequences decreases with the increase of the number of pulses which made a particular Tn sequence. This is naturally accounted for by the fact that as the amount of idle time in the circuit is fixed the increase in the number of pulses of given Tn sequence decreases the amount of Tn sequences that can be inserted. As can be seen from figure~\ref{fig:Grover5qubitsDD} this does not necessarily translate into worse performance for the success probability, where it shows oscillatory behavior with respect to the number of pulses in Tn. This implies that there is interplay or trade-off between the number of inserted DD sequences and the number of pulses in each sequence. The details of the mechanisms at play are further complicated by the different number of DD sequences on each qubit and the calibrations details of every qubit. We have mentioned here briefly qualitatively the factors that influence the results as detailed quantitative study of these mechanisms is beyond the scope of this work. 

\subsection{Success probability for different iterations of Grover operator for chosen dynamical decoupling sequences for the five qubit case}
For completeness of the study we examine the success probability for different iterations of the Grover operator for free evolution, the best performing of CPMG and XY4 and the best performing of topological dynamical decoupling sequences for the given QPU as determined from figure~\ref{fig:Grover5qubitsDD} for five qubit case.

\begin{figure*}[ht]
    \centering
    \includegraphics[width=\linewidth]{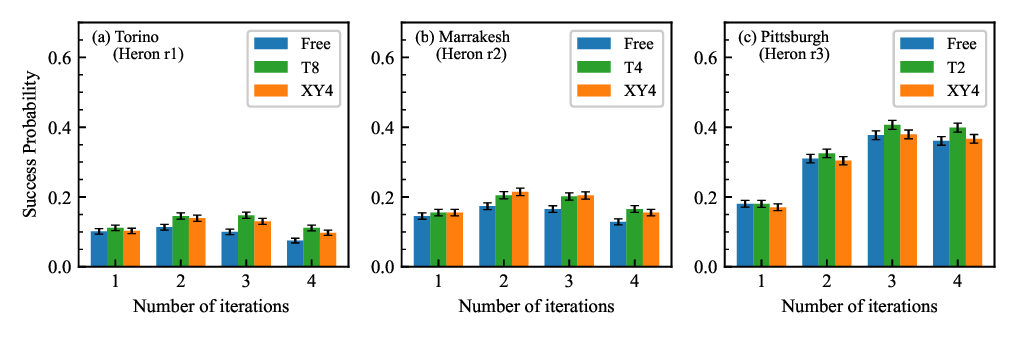}   
    \caption{(Color online)  Success probabilities for Grover's algorithm for five qubits for different number of iterations of Grover operator without dynamical decoupling and with the best dynamical decoupling from the topological sequences and the best of CPMG or XY4 for the given QPU as determined from figure~\ref{fig:Grover5qubitsDD}}
    \label{fig:Grover5qubitsDifferentIterationsWithDD}
\end{figure*}

On figure~\ref{fig:Grover5qubitsDifferentIterationsWithDD} are presented the success probabilities for different number of iterations of Grover operator for the five qubit case on 
(a) Torino (Heron r1), (b) Marrakesh (Heron r2) and (c) Pittsburgh (Heron r3). Previous studies\cite{zhangQuantumSearchNoisy2022,pokharelBetterthanclassicalGroverSearch2024} of five qubit case of Grover's algorithm have focused on up to two iterations due to hardware limitations. Here we consider up to the optimal number of four iterations of the five qubit case. This has been possible due to advancement of quantum computing technology. 

The first observation to note is that the respective Tn sequence offers slightly better enhancement of success probability compared to XY4 for all number of iterations on (a) Torino (Heron r1) and (c) Pittsburgh (Heron r3) while the results on (b) Marrakesh (Heron r2) are of more mixed nature. Second as already noted the improvement of qubit properties and fidelity of gates leads to higher values of the success probabilities for newer generations of QPUs. For both the free case and considered DD sequences the highest success probability is achieved for number of iterations smaller than the theoretically optimal. The considered DD sequences do not alter the iteration for which peak performance is achieved compared to the free case. The highest success probability is achieved for three iterations on Pittsburgh and the measured value of success probability is slightly above 0.4 .

\subsection{Grover's algorithm for six qubit case and different number of iterations of Grover operator} 
Trying to reach the maximum of current day hardware capability we examine the performance of Grover's algorithm for the six qubit case for different iterations of Grover operator for target bitstring "010110" which again as in previous cases is balanced in terms of 0-s and 1-s. We perform the study for free case and circuits with inserted XY4 and T4 dynamical decoupling sequences. XY4 is chosen to CPMG as it consistently shows better enhancement of success probability of the five qubit case compared to CPMG DD sequence. XY4 DD sequence has also consistently performed best compared to CPMG in previous studies\cite{pokharelBetterthanclassicalGroverSearch2024, ezzellDynamicalDecouplingSuperconducting2023}. From the topological DD sequences Tn we choose T4 as it has the same number of pulses as XY4 which offers more adequate base for comparison of the DD sequences. 
	
\begin{figure*}[ht]
    \centering
    \includegraphics[width=\linewidth]{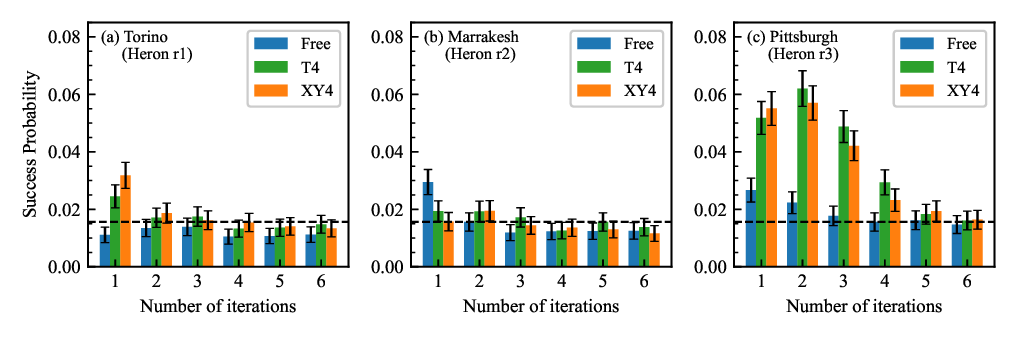}   
    \caption{(Color online)  Success probabilities for Grover's algorithm for six qubits for different number of iterations of Grover operator without dynamical decoupling and with XY4 dynamical decoupling seqeunce and T4 topological dynamical decoupling sequence. XY4 and T4 are compared as they are made of the same number of pulses. The horizontal dashed lines represents the threshold $\frac{1}{2^6}$ for random guess above which we can consider that Grover's algorithm can find the right answer albeit with small probability.}
    \label{fig:Grover6qubitsDifferentIterationsWithDD}
\end{figure*}

The results are presented on figure~\ref{fig:Grover6qubitsDifferentIterationsWithDD} for (a) Torino (Heron r1), (b) Marrakesh (Heron r2) and (c) Pittsburgh (Heron r3) QPUs. Data for mean and min T1 and T2 and mean and max readout error and 2Q gate errors of chosen qubits was gathered several times a day for a period of more than two weeks. The runs for the three platforms were performed at different times chosen when most or all of the mean and min T1 and T2 and mean and max readout error and 2Q gate error were above the mean values of the gathered data. The runs  for (a) Torino were performed on 2026-03-18, for (b) Marrakesh and (c) Pittsburgh the runs were performed on 2026-03-20. The calibration data used for the runs is presented in table~\ref{table:Grover6QubitProperties}. The results for the success probabilities is based on 10000 shots per every run.

The results for the success probability presented on figure~\ref{fig:Grover6qubitsDifferentIterationsWithDD} overall show that in come cases T4 offers better performance enhancement compared to XY4 and in some case XY4 gives better results but overall the two are comparable. For the run on (b) Marrakesh for one iteration both XY4 and T4 actually decrease the success probability compared to the free case, but such behavior is not observed for any other device or number of iterations.

\begin{table}[h]
\caption{Number of 2Q operations used in the calculations of intermediate number of iterations for Grover's algorithm implemented for six qubits.}
\label{table:Grover6IterationsNumberOf2QOperations6Qubit}
\begin{center}
\begin{tabular}{l c c c c c c}
\hline
\diagbox{QC}{Iterations}  & 1  & 2   & 3  & 4  & 5 & 6  \\ 
\hline
Torino (Heron r1) & 375  & 752 & 1117 & 1523 &  1916 &  2315  \\
Marrakesh (Heron r2) & 375  & 752  & 1117 & 1523 & 1916 & 2315  \\
Pittsburgh (Heron r3) & 386  & 765 & 1151 & 1539 &  1908 &  2302 \\
\hline
\end{tabular}
\end{center}
\end{table}

For the case of (a) Torino for one iterations the free case is under the threshold of random guess $\frac{1}{2^6}$ while both the results with T4 and XY4 are above that threshold with XY4 giving better performance. For all other iterations on (a) Torino the free case is under the random guess threshold while the results enhanced with T4 and XY4  are on the random guess threshold within the confidence intervals. On (b) Marrakesh the free case for one iteration is above the random guess threshold, while for one iteration the XY4 and T4 are in the range of the random guess threshold.  For two to six iterations on (b) Marrakesh the success probability of the free, T4 and XY4 cases are slightly below the random guess threshold.

The case of (c) Pittsburgh shows much higher success probabilities with the help of T4 and XY4 for up to three iterations of Grover operator compared to the examined case of (a) Torino and (b) Marrakesh. While the free success probability on (c) Pittsburgh is comparable to the free case on (b) Marrakesh for one iteration when cases with XY4 and T4 DD inserted sequences are considered the result for the success probability for one iterations reaches the value of between 0.05 and 0.055 for T5 and XY4 with XY4 giving slighly better performance compared to T4. For the case of two iterations on (c) Pittsburgh the success probability for T4 reaches above 0.06 and is higher than that of XY4. 

For three iterations on (c) Pittsburgh while the success probability of the free case is in the range of the random threshold value for the six qubit case the values obtained with the help of inserted T4 and XY4 DD sequences is in the rage of 0.04 and 0.05 which is more than two times larger than the random threshold value. The values of the success probabilities with inserted DD sequences decrease with larger number of iterations of Grover operator and approach the free case limit for five and six iterations. It seems that the relative increase achieved with the help of the considered DD sequences is larger for smaller number of iterations which are achieved for shorter circuits. The number of 2Q gates per iteration for the considered QPUs and circuits is presented in table~\ref{table:Grover6IterationsNumberOf2QOperations6Qubit}.

\begin{figure}[ht]
    \centering
    \includegraphics[width=\linewidth]{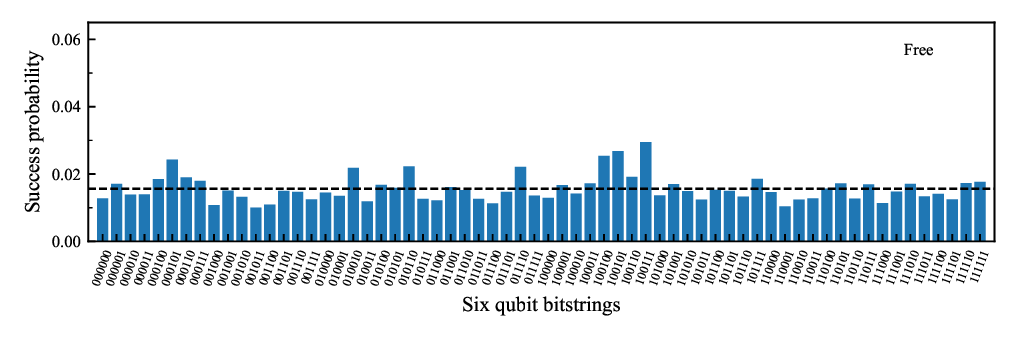}   
    \caption{(Color online)  Success probabilities for different bitstrings for two iterations of Grover operator executed on Pittsburgh (Heron r3) without inserted dynamical decoupling sequence. The horizontal dashed lines represents the threshold $\frac{1}{2^6}$ for random guess.}
    \label{fig:Histogram6QPittsburghTwoIterationsFree}
\end{figure}

\begin{figure}[ht]
    \centering
    \includegraphics[width=\linewidth]{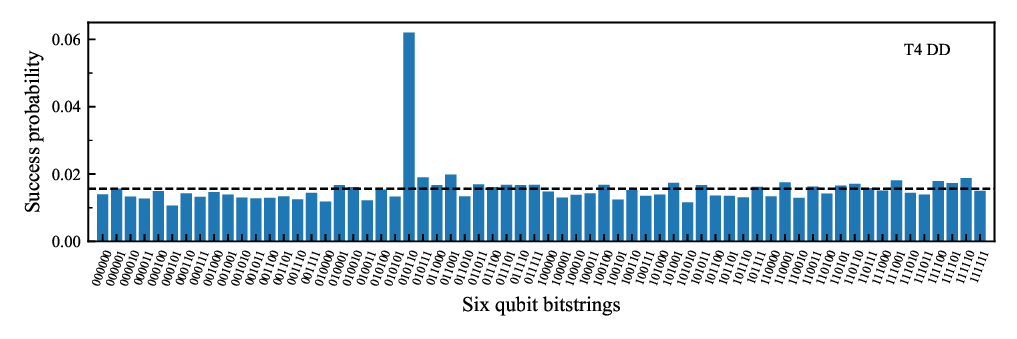}   
    \caption{(Color online)  Success probabilities for different bitstrings for two iterations of Grover operator executed on Pittsburgh (Heron r3) with inserted T4 dynamical decoupling sequence. The horizontal dashed lines represents the threshold $\frac{1}{2^6}$ for random guess.}
    \label{fig:Histogram6QPittsburghTwoIterationsT4}
\end{figure}

\begin{figure}[ht]
    \centering
    \includegraphics[width=\linewidth]{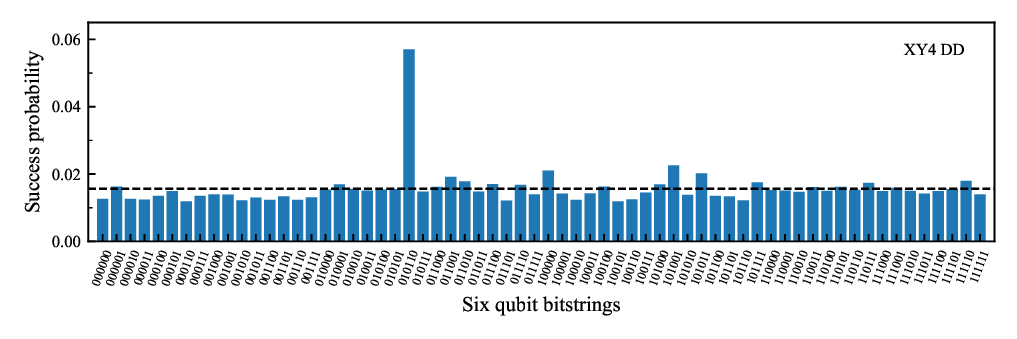}   
    \caption{(Color online) Success probabilities for different bitstrings for two iterations of Grover operator executed on Pittsburgh (Heron r3) with inserted XY4 dynamical decoupling sequence. The horizontal dashed lines represents the threshold $\frac{1}{2^6}$ for random guess.}
    \label{fig:Histogram6QPittsburghTwoIterationsXY4}
\end{figure}

To confirm that the success probabilities for the six qubit case indeed find the target bitstring ``010110'' with the highest probability we plot the success probabilities for all the considered bitstrings for the case of two iterations of Grover operator on Pittsburgh QPU. We choose this case as this offers the highest success probability from the presented on figure~\ref{fig:Grover6qubitsDifferentIterationsWithDD} for the six qubit case. The result for the case without inserted dynamical decoupling is presented on figure~\ref{fig:Histogram6QPittsburghTwoIterationsFree}, the one with inserted T4 DD sequences is presented on figure~\ref{fig:Histogram6QPittsburghTwoIterationsT4} and the case with inserted XY4 DD sequences is presented on figure~\ref{fig:Histogram6QPittsburghTwoIterationsXY4}.

On figure~\ref{fig:Histogram6QPittsburghTwoIterationsFree} for the free case it is evident that the target bitstring  ``010110'' is not the one with the highest probability. Instead there are a few other with higher values of the success probability indicating that the Grover's algorithm does not find the proper sought after solution in this case. In contrast the results with inserted T4 DD sequences presented on  figure~\ref{fig:Histogram6QPittsburghTwoIterationsT4} and the results with inserted XY4 DD sequences presented on  figure~\ref{fig:Histogram6QPittsburghTwoIterationsXY4} both show that the target bitstring  ``010110'' is found with probability much higher than all the other possible bitstrings for the six qubit case. In these two cases the success probabilities of all the other possible bitstrings are around the random threshold value. The result for the success probability achieved with the help of the T4 DD are slightly better than the results achieved with the help of XY4. For the five qubit case the highest value of success probability is achieved for three iterations of Grover operator on Pittsburgh (Heron r3) QPU with the enhancement of T2 DD.

These results show that the combination of advancement of hardware capabilities manifested in better qubit properties with newer generations of QPUs combined with the techniques of dynamical decoupling allow to stretch the Grover's algorithm to work for a six qubit case for reduced number of iterations of Grover operator.

\section{Conclusion}

We have performed selected runs of Grover's algorithm for three, four, five and six qubit cases represented by one bitstring for each case on three different generations of Heron family of superconducting QPUs. The results for three, four and five qubits without dynamical decoupling show higher success probabilities than results reported from studies on previous families of superconducting QPUs even when performed with inserted dynamical decoupling. There is overall improvement of the results within the three generations of Heron family of QPUs with the results for the latest generation showing considerable improvement to all previously considered superconducting QPUs.

For the five qubit bitstring ``01011'' values of success probability were considered for different theoretically suboptimal count of iterations of the Grover operator and the enhancement of the success probability with recently proposed topological dynamical decoupling sequences Tn with between 2 and 12 pulses and compared with the classical CPMG and XY4 DD sequences. On the three considered QPUs the XY4 performs better than CPMG. While the enhancement of XY4 DD and the best of Tn sequences are comparable overall the best Tn sequence for the given QPU gives slightly better results than XY4 DD.

For the six qubit bitstring ``010110'' runs have been performed from one to the theoretically optimum value of six iterations of Grover operator for cases without dynamical decoupling and with T4 DD and XY4 DD on the considered Heron family of superconducting QPUs. The results show that for the theoretically optimal value of iterations of Grover operator the success probability for the target bitstring is around the random threshold value of $\frac{1}{2^6}$ and therefore the Grover's algorithm can not distinguish the sought after solution from all the others in this case. However on Pittsburgh (Heron r3) for reduced number of iterations of Grover operator and with the help of T4 DD or XY4 DD it is possible to distinguish the sought after solution of the six qubit case albeit with very small probability.

%
%

\section*{Acknowledgement}
We acknowledge the use of IBM Quantum services for this work. The views expressed are those of the authors and do not reflect the official policy or position of IBM or the IBM Quantum team.


\section*{Author contributions}
T.G.T. contributed to conceptualization and was responsible for software, visualization and writing - original draft.
N.P.N contributed to conceptualization, software and writing - review \& editing.
N.V.V. was responsible for supervision, conceptualization and writing - review \& editing.



\appendix

\section{Tables with qubit properties for three, four, five and six qubit cases}

\begin{table}[h]
\caption{Qubit properties used for the Grover's algorithm implemented for three qubits}
\label{table:Grover3QubitProperties}
\begin{center}
\begin{tabular}{l  l  l  l  l  l  l  l  l  l}
\hline
Parameter& Torino & Marrakesh & Pittsburgh \\
\hline
Qubits set & [37, 33, 52] & [14, 15, 13] & [119, 133, 113] \\ 
Mean T1 ($\mu s$) & 170.318 & 316.895 & 358.751 \\ 
Min T1 ($\mu s$) & 116.498 & 201.72 & 325.191 \\ 
Mean T2 ($\mu s$) & 142.954 & 122.421 & 389.366 \\ 
Min T2 ($\mu s$) & 61.999 & 28.322 & 377.499 \\ 
Mean Readout Error & 2.071e-02 & 1.367e-02 & 3.052e-03 \\ 
Max Readout Error & 3.992e-02 & 3.027e-02 & 4.395e-03 \\ 
Mean 2Q Error & 1.927e-03 & 1.274e-03 & 8.773e-04 \\ 
Max 2Q Error & 2.044e-03 & 1.306e-03  & 9.591e-04 \\ 
\hline
\end{tabular}
\end{center}
\end{table}

\begin{table}[h]
\caption{Qubit properties used for the Grover's algorithm implemented for four qubits}
\label{table:Grover4QubitProperties}
\begin{center}
\begin{tabular}{l  l  l  l  l  l  l  l  l  l}
\hline
Parameter & Torino & Marrakesh & Pittsburgh \\
\hline
Qubits set & [61, 54, 62, 60] & [11, 18, 12, 10] & [113, 112, 114, 119] \\ 
Mean T1 ($\mu s$) & 174.089 & 199.615 & 301.104 \\ 
Min T1 ($\mu s$) & 106.371 & 117.641 & 226.111 \\ 
Mean T2 ($\mu s$) & 171.981 & 130.375 & 364.685 \\ 
Min T2 ($\mu s$) & 110.561 & 77.212 & 284.282 \\ 
Mean Readout Error & 1.260e-02 & 8.606e-03 & 4.272e-03 \\ 
Max Readout Error & 1.575e-02 & 1.172e-02 & 9.644e-03 \\ 
Mean 2Q Error & 2.563e-03 & 1.743e-03 & 1.111e-03 \\ 
Max 2Q Error & 3.983e-03 & 1.920e-03 & 1.496e-03 \\ 
\hline
\end{tabular}
\end{center}
\end{table}

\begin{table}[h]
\caption{Qubit properties used for the Grover's algorithm implemented for five qubits}
\label{table:Grover5QubitProperties}
\begin{center}
\begin{tabular}{l  l  l  l  l  l  l  l  l  l}
\hline
Parameter & Torino & Marrakesh & Pittsburgh \\
\hline
Qubits set & [65, 55, 64, 66, 46] & [11, 10, 12, 18, 9] & [113, 119, 114, 112, 133] \\ 
Mean T1 ($\mu s$) & 215.307 & 204.577 & 323.201 \\ 
Min T1 ($\mu s$) & 172.005 & 117.641  & 226.111 \\ 
Mean T2 ($\mu s$) & 180.744 & 163.237 & 373.975 \\ 
Min T2 ($\mu s$) & 135.613 & 77.212  & 284.282 \\ 
Mean Readout Error & 2.610e-02 & 1.108e-02  & 4.028e-03 \\ 
Max Readout Error & 6.226e-02 & 2.100e-02 & 7.202e-03 \\ 
Mean 2Q Error & 2.243e-03 & 1.623e-03 & 1.073e-03 \\ 
Max 2Q Error & 3.160e-03 & 1.920e-03 & 1.496e-03 \\ 
\hline
\end{tabular}
\end{center}
\end{table}

\begin{table}[h]
\caption{Qubit properties used for the Grover's algorithm implemented for six qubits}
\label{table:Grover6QubitProperties}
\begin{center}
\begin{tabular}{l  l  l  l  l  l  l  l  l  l}
\hline
Parameter & Torino & Marrakesh & Pittsburgh \\
\hline
Qubits set & [65, 55, 66, 64, 46, 67] & [3, 2, 4, 16, 1, 5] & [85, 86, 84, 77, 87, 88] \\ 
Mean T1 ($\mu s$) & 189.624 & 221.871 & 338.628 \\ 
Min T1 ($\mu s$) & 80.533 & 137.07  & 202.779 \\ 
Mean T2 ($\mu s$) & 169.534 & 204.839 & 359.942 \\ 
Min T2 ($\mu s$) & 100.681 & 75.636  & 257.751 \\ 
Mean Readout Error & 2.840e-02 & 5.981e-03  & 3.581e-03 \\ 
Max Readout Error & 5.615e-02 & 8.301e-03 & 5.371e-03 \\ 
Mean 2Q Error & 2.264e-03 & 1.826e-03 & 1.103e-03 \\ 
Max 2Q Error & 3.014e-03 & 1.886e-03 & 1.492e-03 \\ 
\hline
\end{tabular}
\end{center}
\end{table}

\clearpage

\printbibliography

\end{document}